\documentclass[12pt]{article}

\usepackage[margin=1in,footskip=0.25in]{geometry}
\usepackage{amsthm}%
\usepackage{mathtools}
\usepackage{amsfonts}%
\usepackage{pdfpages}
\usepackage{amsmath}
\usepackage{color}

\usepackage{cite}
\newtheorem{theorem}{Theorem}

\newtheorem{definition}[theorem]{Definition}

\begin{document}

\title{Transient hidden chaotic attractors in a Hopfield neural system}

\author{Marius-F. Danca{\footnote{Corresponding author}}\\
Department of Mathematics and Computer Science, \\Avram Iancu University,\\
400380 Cluj-Napoca, Romania,\\and\\Romanian Institute of Science and Technology, \\
400487 Cluj-Napoca, Romania\\
Email: danca@rist.ro\\
\vspace{3mm}\\
Nikolay Kuznetsov\\
Department of Applied Cybernetics, \\Saint-Petersburg State University, Russia\\
and\\
Department of Mathematical Information Technology, \\University of Jyv\"{a}skyl\"{a}, Finland\\
email: nkuznetsov239@gmail.com}

\maketitle

\begin{abstract}In this letter we unveil the existence of transient hidden coexisting chaotic attractors, in a simplified Hopfield neural network with three neurons.
\end{abstract}

\textbf{keyword }Hopfield neural network; Transient hidden chaotic attractor; Limit cycle

\vspace{3mm}

\section{Introduction}

A Neural Network (NN) is a mathematical or computational model inspired by biological neural networks that consists of interconnected groups of neurons.
Without chaotic behavior neural systems cannot be adequately addressed and fully understood \cite{skarda}. Neurobiological chaos, omnipresent in the brain, points out several possible approaches of understanding how the brain works and this is demonstrably so, in the somatosensory and the olfactory cortices \cite{wal}.
Many NNs, such as discrete time NNs, or continuous (time-delayed) NNS, may behave chaotically.

The roles of chaos in this type of systems have been investigated in many papers in the last years \cite{gid,nara2,cao,aih,free,guck}.

Hopfield Neural Networks (HNN) are constructed from artificial neurons and represent particular cases of NNs inspired by spin systems. Even if it is not easy to be discovered, chaos and hyperchaos have been identified in many HNNs \cite{hop,hug,ber,li,yang,yangy,yangx,huang,yuoan,xsyang,wzhuang,qli,alo,cpb4,cpb5,cpb6,cpb7}.

On the other side, the origin of transient chaos is well known: it is due to nonattracting chaotic saddles in phase space \cite{ttr1,ttr2,tr1,tr2,tutu2,tr3,tr4,cpb2,tr5,tr6,tutu1,cpb3}. Transient chaos is a common phenomenon of many engineering, physical and biological systems.
Compared with chaos, which is characterized as a long-term behavior, the transient chaos, is a phenomenon which appears when a nonlinear system behaves chaotically during some transient time interval which falls after that into a periodic attractor. These systems may initially exhibit an aperiodic behavior and sensitivity to initial conditions (i.e. ``chaos'') and after a period of time, it settles down on a periodic orbit or fixed point. Such phenomena were observed in radio circuits \cite{treix}, hydrodynamics \cite{doix}, neural networks \cite{patrux}, standard models of nonlinear systems such as R\"{o}ssler system \cite{cincix}, Lorenz system \cite{sasex}, experiments \cite{zerox}, maps \cite{asta}, species extinction \cite{tutu3} and so on.

In some applications the transient chaos can be quite disastrous, as in situations of voltage collapse or species extinction. Therefore it is often desirable to sustain transient chaos, in the sense of maintaining the transient chaos. Thus, conversion of the transient chaos into sustained chaos, can avoid catastrophes related to sudden chaos collapses, even in the absence of external perturbations.  (i.e. chaos anticontrol) \cite{ttrx,tutu2}.

In a recent paper \cite{mot} a new phenomenon of transient chaos, fundamentally different from the hyperbolic and nonhyperbolic transient chaos reported in the existing literature is revealed. This type of phenomenon appears in many systems (chemical reactions, binary star behavior, etc.) and it is likely far less predictable than has been previously thought: \emph{doubly transient chaos}.

From the computational perspective point of view, based on the connection of their basins of attraction with equilibria in the phase space, it is natural to suggest the following attractors classification \cite{ixus3,unu,doi,patru,optimus}

\begin{definition}\label{def}
An attractor is called a \emph{self-excited attractor} if its basin of attraction intersects
with any open neighborhood of a stationary state (an equilibrium); otherwise, it is called a \emph{hidden attractor}.
\end{definition}

Self-excited attractors can be visualized numerically by a standard computational procedure, in which
after a transient process, a trajectory starting from a point of a neighborhood of unstable equilibrium
is attracted to the attractor, while the basin of attraction for a hidden attractor is not connected with
any equilibrium. Therefore, for the numerical localization of hidden attractors it is necessary to develop
special analytical-numerical procedures \cite{unu,trei}.

Hidden attractors can appear in systems with no-equilibria or in multistable systems with only stable equilibrium.
Coexisting self-excited attractors in multistable systems can
be found using a standard computational procedure, whereas there is no regular way to predict the existence
or coexistence of hidden attractors in a system (for various examples of multistable engineering systems refer to \cite{opt}).

To verify numerically that a chaotic attractor is hidden, one has to check that all trajectories starting in small neighborhoods of unstable equilibria, are either attracted by stable attractors, or diverge to infinity (see e.g. \cite{doi,trei}).

In this paper we consider the case of a $3$-neuron simplified HNN and we unveil its transient hidden chaotic attractors.

\section{The simplified Hopfield neural network}

The simplified $3$-neuron HNN considered in this paper is modeled by the following ODEs

\begin{equation}\label{eq}
\dot{x}_i=-x_i+\sum_{j=1}^3w_{ij}f(x_j),\quad i=1,2,3,
\end{equation}

\noindent with $f(x_j)=tanh(x_j)$ and with the weight matrix \cite{yang}

\[W=
\begin{bmatrix}
    w_{11}& w_{12}& w_{13}\\
    w_{21}& w_{22}& w_{23}\\
    w_{31}& w_{32}& w_{33}
\end{bmatrix}
=
\begin{bmatrix}
2&-1.2&0\\
1.9995& 1.71&1.15\\
-4.75&0&1.1
\end{bmatrix}.
\]

\noindent and we show numerically that the system admits a new type of coexisting hidden attractors: \emph{transient hidden attractors}\footnote{Transient dynamics of hidden attractors in a 4D system are analyzed in \cite{cpb1}.}.

With the above values of weights, the system (\ref{eq}) reads
\begin{equation}\label{eq2}
\begin{aligned}
\dot{x}_1=&-x_1+2~tanh(x_1)-1.2~tanh(x_2),\\
\dot{x}_2=&-x_2+1.9995~tanh(x_1)+1.71~tanh(x_2)+1.15~tanh(x_3),\\
\dot{x}_3=&-x_3-4.75~tanh(x_1)+1.1~tanh(x_3).
\end{aligned}
\end{equation}

The sigmoid-like function $tanh(x)$, is used to approximate the switch discontinuity in $x=0$, typically to neurons dynamics.

The HNN system (\ref{eq2}) is symmetrical with respect to the origin and has the following equilibria

\[
X_0^*=(0,0,0),\quad X_{1,2}^*=\pm(0.493,0.366,-3.267).
\]

The Jacobian is
\[J=
\begin{bmatrix}
    1-2~tanh^2(x_1)&-1.2+1.2~tanh^2(x_2)&0\\
1.9995-1.9995~tanh^2(x_1)&      0.71-1.71~tanh^2(x_2)&1.15-1.15~tanh^2(x_3)\\
-4.75+4.75~tanh^2(x_1)&0&0.1-1.1~tanh^2(x_3)
\end{bmatrix},
\]

\noindent and the eigenvalues of $X_0^*$ are $\lambda_1=1.942$ and $\lambda_{2,3}=-0.066 \pm 1.879i$ while the eigenvalues of $X_{1,2}^*$ are $\lambda_1=-0.987$ and $\lambda_{2,3}=0.538 \pm 1.286i$. Therefore, equilibria are unstable: one attracting focus saddle ($X_0^*$), and two repelling focus saddles ($X_{1,2}^*$).

\section{Hidden transient attractors of HNN}

 The numerical integration of the HNN (\ref{eq2}) is done with the Matlab differential solver $ode45$ with option  $opts = odeset('RelTol',1e-9,'AbsTol',1e-9)$ which yields $8$ decimals accurate results, over the time interval $[0,T]$ with $T=850$.\footnote{Matlab implicitly uses default values $RelTol = .001$ and $AbsTol = .000001$ and the approximate error at each step $e_k$ is ensured to be: $e_k\leq max(RelTol\times x_k, AbsTol)$, for all $k$, where $x_k$ is the value calculated at the node $t_k$ (see e.g. \cite{num2} for the used Runge-Kutta and other numerical methods utilized by Matlab). In \cite{num3} it is suggested $RelTol=10^{-(m+1)}$ for $m$ precise digits of the required solution.}

In Hopfield like systems it is common to find transient chaos (see e.g. \cite{yang}). Generally, the duration of a chaotic transient behavior of many trajectories is rather short before they settle down on some periodic stable attractor \cite{tr4}. However, in the case of the HNN, our various numerical tests reveal a relative long transient time $[0,T^*]$, with $T^*=500$, over which the system behaves chaotically along two coexisting transient chaotic attractors $H_1$ and $H_2$ (see Fig.\ref{fig3} where $H_{1,2}$ are obtained with initial conditions $\pm(1.9,3,1)$)\footnote{Related to length of the time interval, precautions should be considered since a too large time interval could lead to inaccurate numerical solutions (see e.g. \cite{long} for the case of Lorenz system).}.

For the initial conditions chosen inside small $\delta$-vicinities of the unstable equilibria, with $\delta=1.5E-4$, the underlying trajectories are attracted by one of the two stable limit cycles $C_1$ (red plot) and $C_2$ (blue plot) respectively as shown by Figs. \ref{fig4} and \ref{fig5} where, for simplicity, in the detailed images only 50 trajectories are considered.

Trajectories starting from a $\delta$-vicinity $V_{X_0^*}$ of $X_0^*$ (Fig. \ref{fig4}), tend either to $C_1$ (light red plot), or to $C_2$ (light blue plot). This happens because $X_0^*$ belongs to the separatrix of the basins of attractions of $C_1$ and $C_2$.

Fig. \ref{fig5}, where only the case of stable limit cycle $C_2$ is considered, reveals 50 trajectories starting inside a vicinity $V_{X_2^*}$ of equilibrium $X_2$, which all are attracted by the stable cycle $C_2$. Similarly, all trajectories from small vicinities of $X_1^*$ are attracted by $C_1$.

The shape of the trajectories starting within $V_{X_0^*}$ and $V_{X_{1,2}^*}$ are consistent with the equilibria type: the trajectories from vicinities $V_{X_{1,2}^*}$ exit by scrolling equilibria $X_{1,2}$ in the unstable two-dimensional manifold (see detailed image in Fig. \ref{fig4}), while the trajectories from the vicinity of $X_0^*$ leave $V_{X_0^*}$ along the one-dimensional unstable manifold of $X_0^*$ (detail in Fig. \ref{fig5}).

Concluding, by following intensive numerical tests, the underlying numerical analysis leads to the conclusion that the transient chaotic attractors $H_{1,2}$, are very likely to be hidden and could be called \emph{transient hidden chaotic attractors}. For the numerical localization of hidden attractors, a special analytical-numerical procedure has been designed (see e.g. \cite{trei}). In this paper the initial conditions for transient hidden attractors $\pm(1.9,3,1)$, have been found by trial-and-error.

Note that the coexistence of transient trajectories which start from $V_{X_0^*}$ and $V_{X_{1,2}^*}$ and reach the stable cycles $C_{1,2}$ with the stable cycles $C_{1,2}$, are ensured by the entrainment of limit cycles by chaos (see \cite{theo}, where the replication of sensitivity and the existence of infinitely many unstable periodic solutions were rigorously proved and \cite{theo2}, where this result is applied in Hopfield systems). Based on this result, the transient hidden chaotic attractors $H_{1,2}^*$ differ from the transients to $C_{1,2}$. Moreover, they have different attraction basins: $H_{1,2}$ are generated starting from initial conditions $\pm(1.9,3,1)$, while $C_{1,2}$ are obtained with initial points close to equilibria $X_0^*$ and $X_{1,2}^*$. For $t>T^*$, $T_{1,2}$ vanish.

The transients of the HNN system seems to be deformed by the form of the coexisting limit cycles and unstable equilibria, thus $H_1$ and $H_2$ have a complex structure. So one may say that the behavior is rather natural here.

Since chaotic behavior in neural activity seems to be unavoidable, chaos control and anticontrol of these transient hidden chaotic attractors are an unexplored theme yet and they offer an exciting subject for future research.

\newpage{\pagestyle{empty}\cleardoublepage}

\newpage{\pagestyle{empty}\cleardoublepage}

\begin{figure}
\begin{center}
\includegraphics[scale=0.8]{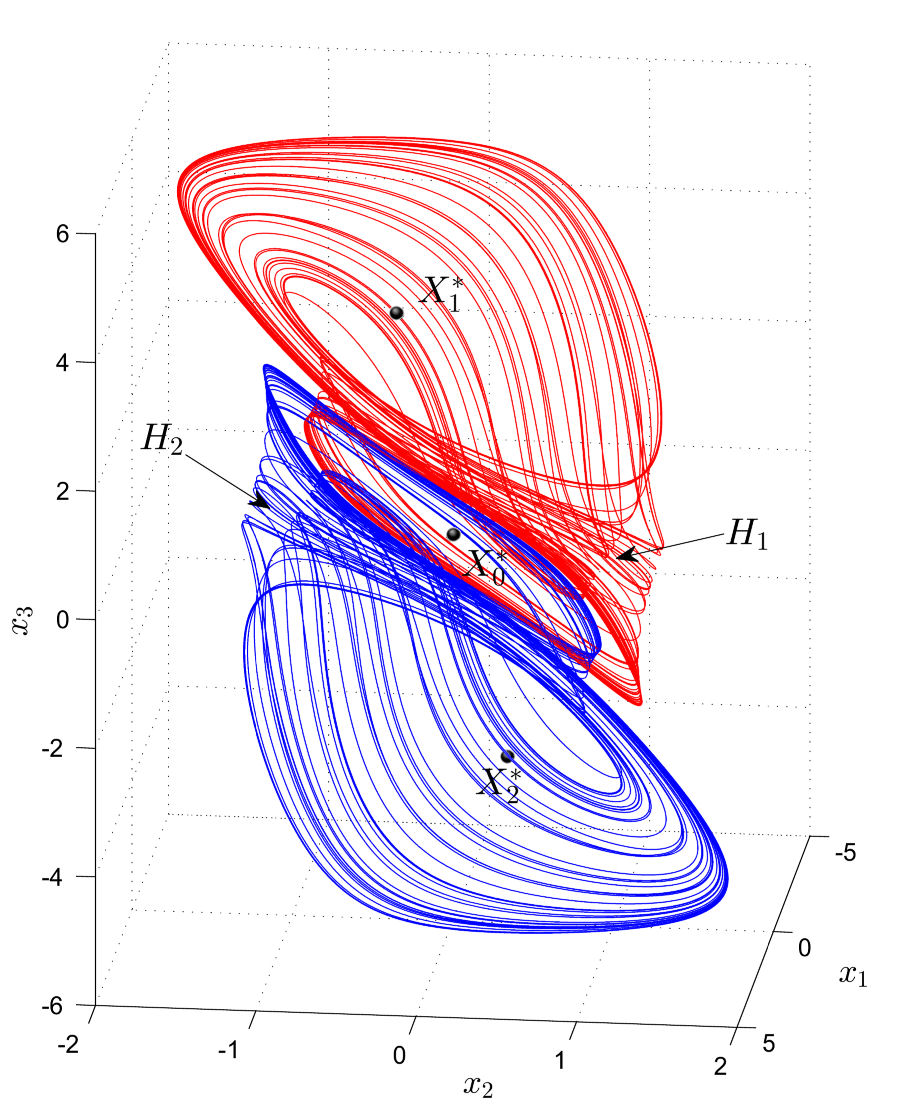}
\caption{Transient hidden chaotic attractors $H_1$ and $H_2$ of HNN.}
\label{fig2}
\end{center}
\end{figure}

\begin{figure}
\begin{center}
\includegraphics[scale=0.8]{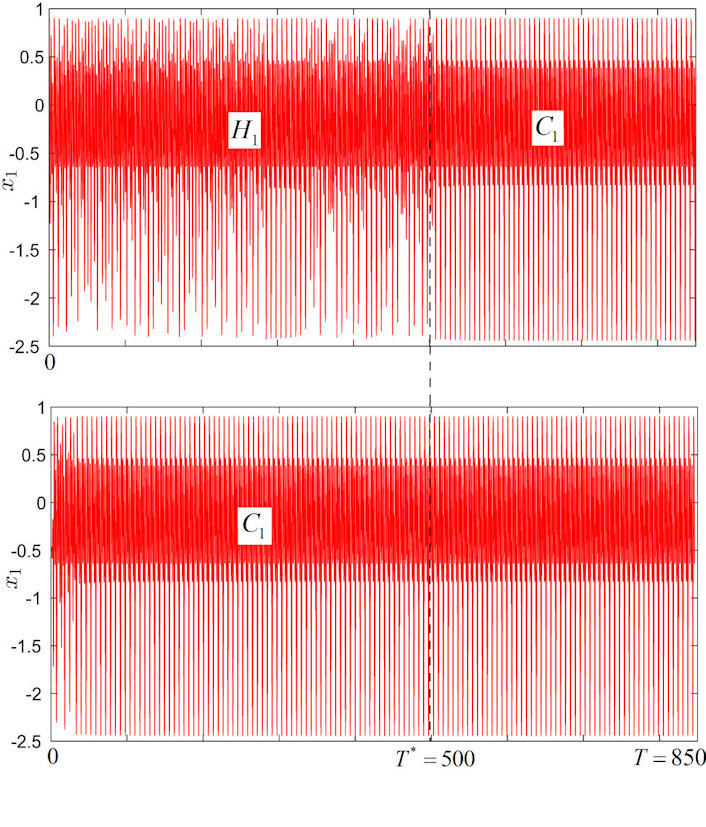}
\caption{Time series of the first component $x_1$. (a) Transient hidden chaotic attractor $H_1$, for $t\in[0,T^*]$, starting from $(1.9,3,1)$. For $t>T^*$, chaos vanishes. (b) Stable cycle $C_1$.}
\label{fig3}
\end{center}
\end{figure}

\begin{figure}
\begin{center}
\includegraphics[scale=0.8]{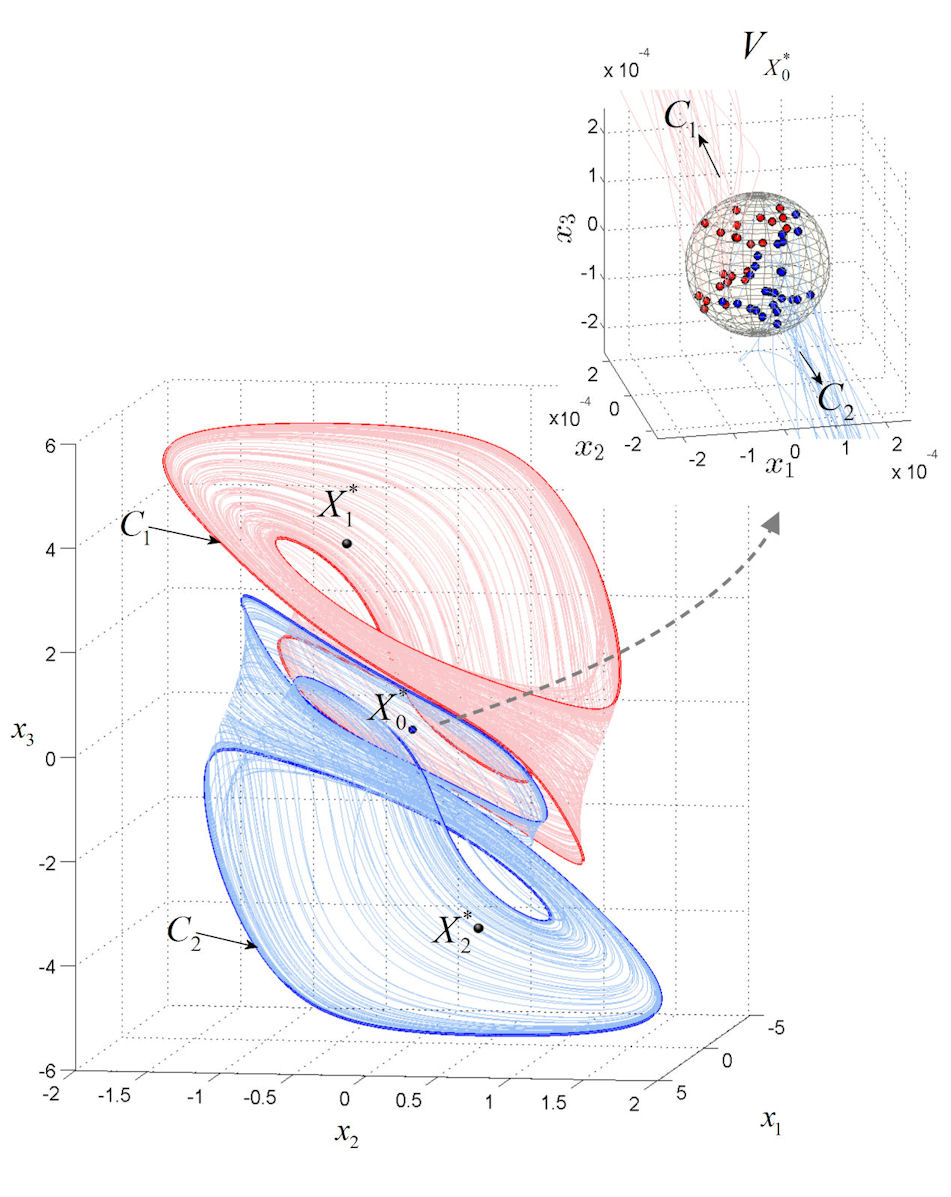}
\caption{Stable cycles $C_1$ (red plot) and $C_2$ (blue plot). Chaotic trajectories $H_{1,2}$ (light red and light blue plots) starting from vicinity $V_{X_0^*}$, are attracted either by $C_1$, or $C_2$ (see also the upper-right detail).}
\label{fig4}
\end{center}
\end{figure}

\begin{figure}
\begin{center}
\includegraphics[scale=0.8]{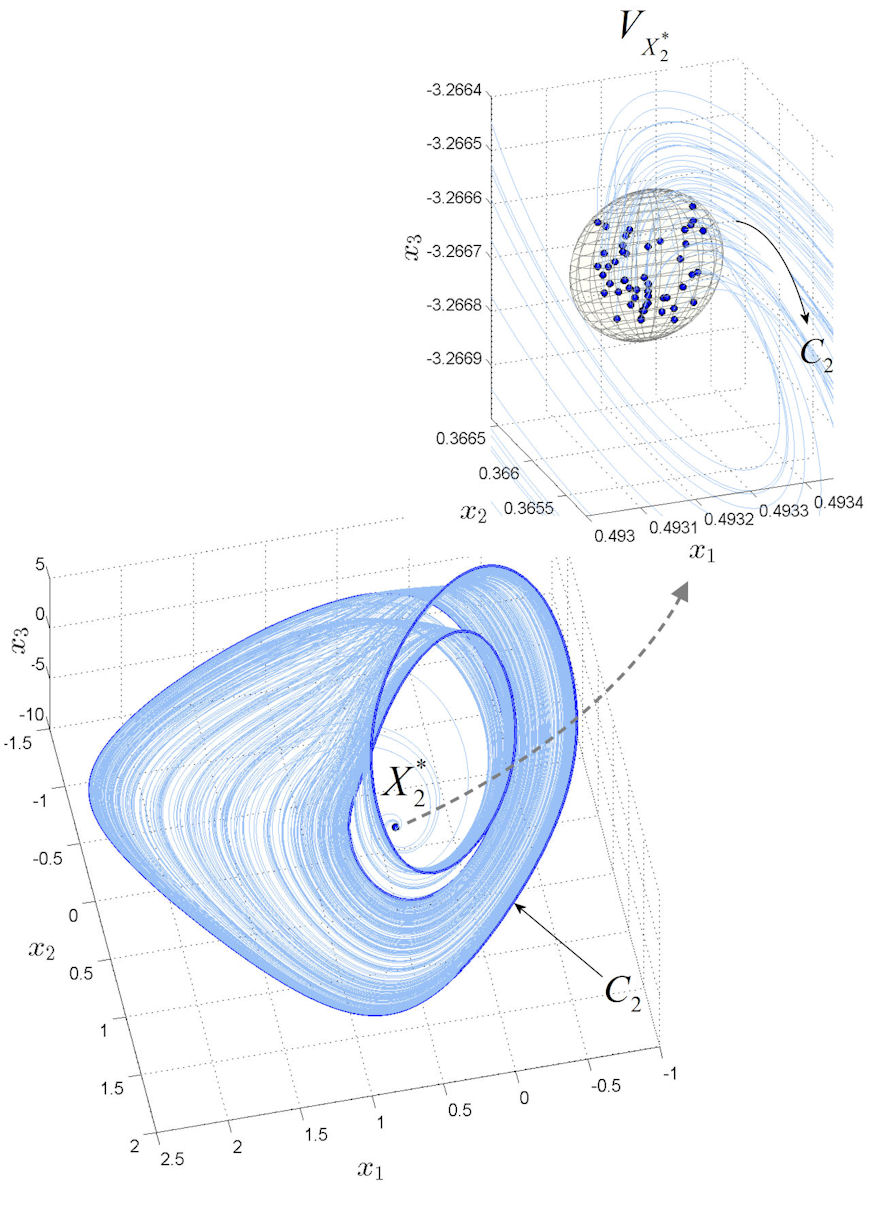}
\caption{Trajectroies starting from vicinity $V_{X_2^*}$ (light blue plot) are attracted by $C_2$ (blue plot).}
\label{fig5}
\end{center}
\end{figure}


\begin{thebibliography}{100}

	\bibitem{skarda} Skarda C A and Freeman W J 1987 \emph{Behav. Brain Sci.} \textbf{10} 161

	\bibitem{wal} Freeman W J 2000 Neurodynamics: An exploration in mesoscopic brain dynamics (Berlin, Springer)

	\bibitem{gid} Dror G and Tsodyks M 2000 \emph{Neurocomputing} \textbf{32-33} 365

	\bibitem{nara2} Nara S, Davis P, Kawachi M and Totsuji H 1995 \emph{Int. J. Bifurcat. Chaos} \textbf{5} 1205

	\bibitem{cao} Cao J and Lu J 2006 \emph{Chaos} \textbf{16} 013133

	\bibitem{aih} Aihara K, Takebe T and Toyoda M 1990 \emph{Phys. Lett. A} \textbf{144} 333

	\bibitem{free} Freeman W J 1992 \emph{Int. J. Bifurcat. Chaos} \textbf{2} 451

	\bibitem{guck} Guckenheimer J and Oliva R A 2002 \emph{SIAM J. Appl. Dyn. Syst.} \textbf{1} 105

	\bibitem{hop} Hopfield J J 1994 \emph{Proc. Nat. Acad. Sci. USA} \textbf{81} 3088

	\bibitem{hug} Bersini H and Sener P 2002 \emph{Neural Networks} \textbf{15} 1197

	\bibitem{ber} Bersini H 1998 \emph{Neural Networks} \emph{11} 1017

	\bibitem{li} Li Q, Yang X S and Yang F 2005 \emph{Neurocomputing} \textbf{67} 275

	\bibitem{yang} Yang X S and Yuan Q 2005 \emph{Neurocomputing} \textbf{69} 232

	\bibitem{yangy}  Chen P-F, Chen Z-Q and Wu W-J 2010 \emph{Chinese Phys B} \textbf{19} 040509

	\bibitem{yangx} Yang X S and Huang Y 2006 \emph{Chaos} \textbf{16} 033114

	\bibitem{huang} Huang W-Z and Huang Y 2011 \emph{Int. J. Bifurcat. Chaos} \textbf{21} 885

	\bibitem{yuoan} Yuan Q, Li Q and Yang X-S 2009 Chaos Solitons Fractals 39 1522

	\bibitem{xsyang} Yang X-S and Yuan Q 2005 \emph{Neurocomputing} \textbf{69} 232

	\bibitem{wzhuang} Huang W-Z and  Huang Y 2008 \emph{Appl. Math. Comput.} \textbf{206} 1

	\bibitem{qli} Li Q, Yang X-S and Yang F 2005 \emph{Neurocomputing} \textbf{67} 275

	\bibitem{alo} Alonso H, Mendon T and Rocha P 2009 \emph{Neural Netw.} \textbf{22} 450

	\bibitem{cpb4} Peng-Sheng Z, Wan-Sheng T, and Jian-Xiong Z 2010 \emph{Chinese Phys B} \textbf{19} 030514

	\bibitem{cpb5} Yi-Fu F, Qing-Ling Z, and De-Zhi F 2012 \emph{Chinese Phys B} \textbf{21} 100701

	\bibitem{cpb6} Park M J and Kwon O M 2011 \emph{Chinese Phys B }\textbf{45} 45013425
\bibitem{cpb7} Vasovi\'{c} N, Buri\'{c} N, Todorovi\'{c} K, and Grozdanovi\'{c} I 2012 \emph{Chinese Phys. B} \textbf{21}, 010203.

	\bibitem{ttr1} Grebogi C, Ott E and Yorke J A 1983 \emph{Phys. Rev. Lett.} \textbf{48 }1507

	\bibitem{ttr2} Kantz H and Grassberger P 1985 \emph{Physica D} \textbf{17} 75

	\bibitem{tr1} Kaplan J L and Yorke J A 1979 \emph{Commun. Math. Phys.} \textbf{67} 93

	\bibitem{tr2} Yorke J A and Yorke E D 1979 \emph{J. Stat. Phys.} \textbf{21} 263

	\bibitem{tutu2} Shulenburger L, Lai Y-C, Yal\c{c}inkaya T and Holt R D 1999 \emph{Phys. Lett. A} \textbf{260} 156

	\bibitem{tr3} Pianigiani G and Yorke J A 1979 \emph{Trans. Am. Math. Soc.} \textbf{252} 351

	\bibitem{tr4} Nusse H E and Yorke J A 1989 \emph{Physica D} \textbf{36} 137

	\bibitem{cpb2} Bo-Cheng B, Zhong L and Jian-Ping X 2010 \emph{Chinese Phys B}  \textbf{19} 030510

	\bibitem{tr5} Hsu G-H, Ott E and Grebogi C 1988 \emph{Phys. Lett. A }\textbf{127} 199

	\bibitem{tr6} Grebogi C, Ott E and Yorke J A1986 \emph{Phys. Rev. Lett.} \textbf{57} 1284

	\bibitem{tutu1} Hoff A, da Silva D T, Manchein C and Albuquerque H A 2014 \emph{Phys. Lett. A} \textbf{378} 171

	\bibitem{cpb3} Chang-Chun S, Qi-Cheng X and Ying S 2013 \emph{Chinese Phys B} \textbf{22} 030507

	\bibitem{treix} Zhu L, Raghu A and Lai Y-C 2001 \emph{Phys.  Rev.  Lett. }\textbf{86} 4017

	\bibitem{doix} Ahlers G, Walden R W 1980 \emph{Phys.  Rev.  Lett. }\textbf{44} 445

	\bibitem{patrux} Yang X-S, Yuan Q 2005 \emph{Neurocomputing} \textbf{69} 232

	\bibitem{cincix} Dhamala M, Lai Y-C and Kostelich E J 2000 \emph{Phys. Rev. E }\textbf{61} 6485

	\bibitem{sasex} Vadasz P 2010 \emph{Appl. Math. Lett.} \textbf{23} 503

	\bibitem{zerox} Zhu L, Raghu A, and Lai Y-C 2001 \emph{Phys. Rev. Lett. }\textbf{86} 4017

	\bibitem{asta} Astaf'ev G B, Koronovski\i{} A A and Hramov A E 2003 \emph{Tech. Phys. Lett.}+ \textbf{29} 923

	\bibitem{tutu3} McCann K and Yodzis P 1994 \emph{Am. Nat.} \textbf{144} 873

	\bibitem{ttrx} Dhamala M. and Lai Y-C 1999 \emph{Phys. Rev. E} \textbf{59} 1646

	\bibitem{mot} Motter A E, Gruiz M, K\'{a}rolyi G and T\'{e}l T 2013 \emph{Phys. Rev. Lett. }\textbf{111} 194101

	\bibitem{ixus3} G. Bianchi, N.V. Kuznetsov, G.A. Leonov, M.V. Yuldashev, R.V. Yuldashev, 2015 7th International Congress on Ultra Modern Telecommunications and Control Systems and Workshops (ICUMT), Limitations of {PLL} simulation: hidden oscillations in MATLAB and SPICE, 79-84 (2015) http://arxiv.org/pdf/1506.02484.pdf,  http://www.mathworks.com/matlabcentral/fileexchange/52419-hidden-oscillations-in-pll

	\bibitem{unu} Leonov G A and Kuznetsov N V 2013 \emph{Int. J. Bifurcat. Chaos} \textbf{23} 1330002

	\bibitem{doi} eonov G, Kuznetsov N. and Mokaev T 2015 \emph{Eur. Phys. J. Special Topics} \textbf{224} 1421

	\bibitem{patru} Leonov G A, Kuznetsov N V and Vagaitsev V I 2011 \emph{Phys. Lett. A} \textbf{375} 2230

	\bibitem{optimus} Sprott J C, Jafari S, Pham V-T and Hosseini Z S 2015 \emph{Phys. Lett. A} \textbf{379} 2030

	\bibitem{trei} Leonov G A, Kuznetsov N V and Vagaitsev V I 2012 \emph{Physica D} \textbf{241} 1482

	\bibitem{opt} Pisarchik A and Feudel U 2014 \emph{Physics Reports }\textbf{540} 167

	\bibitem{cpb1} Xiao-Yu D, Chun-Biao L, Bo-Cheng B and Hua-Gan W 2015 \emph{Chinese Phys. B} \textbf{24} 050503

	\bibitem{num2} Shampine L F, Gladwell I and Thompson S 2003 \emph{Solving ODEs with MATLAB Cambridge} (Univ. Press, Cambridge)

	\bibitem{num3} Brenan K E, Campbell S L and Petzold L R 1996 \emph{Numerical Solution  of Initial-Value Problems in Differential-Algebraic Equations} (SIAM Classics in Applied Mathematics, 14) (SIAM, Philadelphia)

	\bibitem{long} Kehlet B. Logg A arXiv:1306.2782v1 [math.NA] 12 Jun 2013 http://arxiv.org/abs/1306.2782

	\bibitem{theo} Akhmet M U and Fen M O 2014 \emph{J. Nonlinear Sci.} \textbf{24} 411

	\bibitem{theo2} Akhmet M and Fen M O 2014 \emph{Neurocomputing} \textbf{145} 230

	\bibitem{nous} Chaudhuri U and Prasad A 2014 \emph{Phys. Lett. A} \textbf{378} 713


\end{thebibliography}
\end{document}